\documentclass[
superscriptaddress,
amsmath,amssymb,
aps,
twocolumn,
pra,
]{revtex4-1}

\usepackage{pdftricks}
\begin{psinputs}
	\usepackage{pstricks}
	\usepackage{multido}
\end{psinputs}
\usepackage{enumitem}

\usepackage{graphicx}% Include figure files
\usepackage{dcolumn}% Align table columns on decimal point
\usepackage{bm}% bold math
\usepackage{blindtext}
\newcommand{\R}{\mathbb{R}}
\newcommand{\A}{\mathcal{A}}
\newcommand{\B}{\mathcal{B}}
\newcommand{\C}{\mathcal{C}}
\newcommand{\norm}[1]{\lvert #1 \rvert}

\usepackage{color}

\makeatother
\graphicspath{ {figs/} }

\numberwithin{equation}{section}

\begin{document}

	\title{On the geometric phenomenology of static friction}

	\author{Shankar Ghosh$ ^{1} $, A. P. Merin}
	\affiliation{Department of Condensed Matter Physics and Materials Science, }%Lines break automatically or can be forced with \\
	%\author{}
	% \affiliation{Department of Condensed Matter Physics and Materials Science, Tata Institute of Fundamental Research, Mumbai 400005, India}%Lines break automatically or can be forced with \\

	\author{Nitin Nitsure}%
	\affiliation{School of Mathematics, Tata Institute of Fundamental Research, Mumbai 400005, India}
	
	%\date{\today}% It is always \today, today,
	%  but any date may be explicitly specified
		\begin{abstract}
        In this note we introduce a hierarchy of phase spaces for static
		friction, which give a graphical way to systematically quantify the directional dependence in static friction via subregions of the phase spaces. We experimentally plot these subregions 
		to obtain phenomenological descriptions for static friction in various examples
		where the macroscopic shape of the object affects the frictional response.
		The phase spaces have the universal property that for any experiment
		in which a given object is put on a substrate fashioned from a chosen material with a specified nature of contact, 
		the frictional behaviour can be read off from a uniquely determined classifying map on the control space of the experiment which takes values in the appropriate phase space.
		\end{abstract}
	
	\pacs{Valid PACS appear here}
	% PACS, the Physics and Astronomy
	% Classification Scheme.
	%\keywords{Suggested keywords}
	%Use showkeys class option if keyword
	%display desired
	\maketitle
	
	\begin{figure}[b]
\includegraphics[width=1\linewidth]{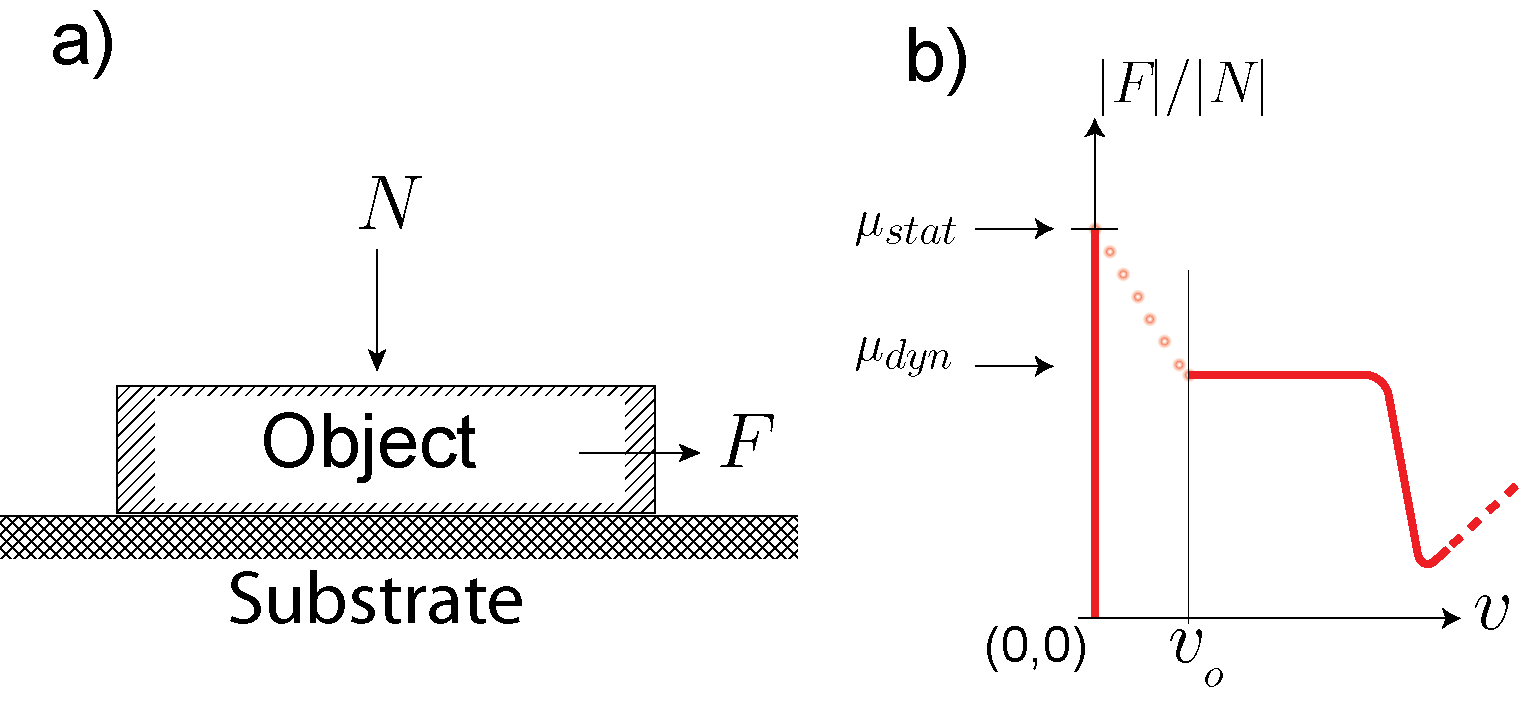}
\caption{(a) A  schematic representation of the experimental setup to measure friction between an object and the substrate where $N$ is the applied normal force and $F$ is the applied tangential force. The point of application of $F$ is kept low enough so as to minimise the toppling torques. 
(b) The red curve consists of all possible pairs $(v,|F|/|N|)$ for which the object can be driven at
a fixed velocity $v$ for a moderate $|N|$. A static object ($v=0$) remains 
stationary if $|F|<\mu_{stat} |N|$. 
The force of friction drops as the object starts to move, and its magnitude 
decreases from higher values to a lower steady value $\mu_{dyn} |N|$ for a certain range of velocities,
beginning with a velocity $v_0$.  
The resulting curve is flat in a small interval starting from $v=v_0$, 
where the value of  $|F|/|N|$ is  $\mu _{dyn}$. The exact shape of the curve between $0$ and $v$ is not known as it is a region of intermittent response.}
% however, in the interval $0< v<v_0$  the value of $|F|/|N|$ is strictly greater than $\mu_{dyn}$.} 
\label{fig:velo_gap}
	\end{figure}
	
	\begin{figure}[b]
		\centering
		\includegraphics[width=0.45\linewidth]{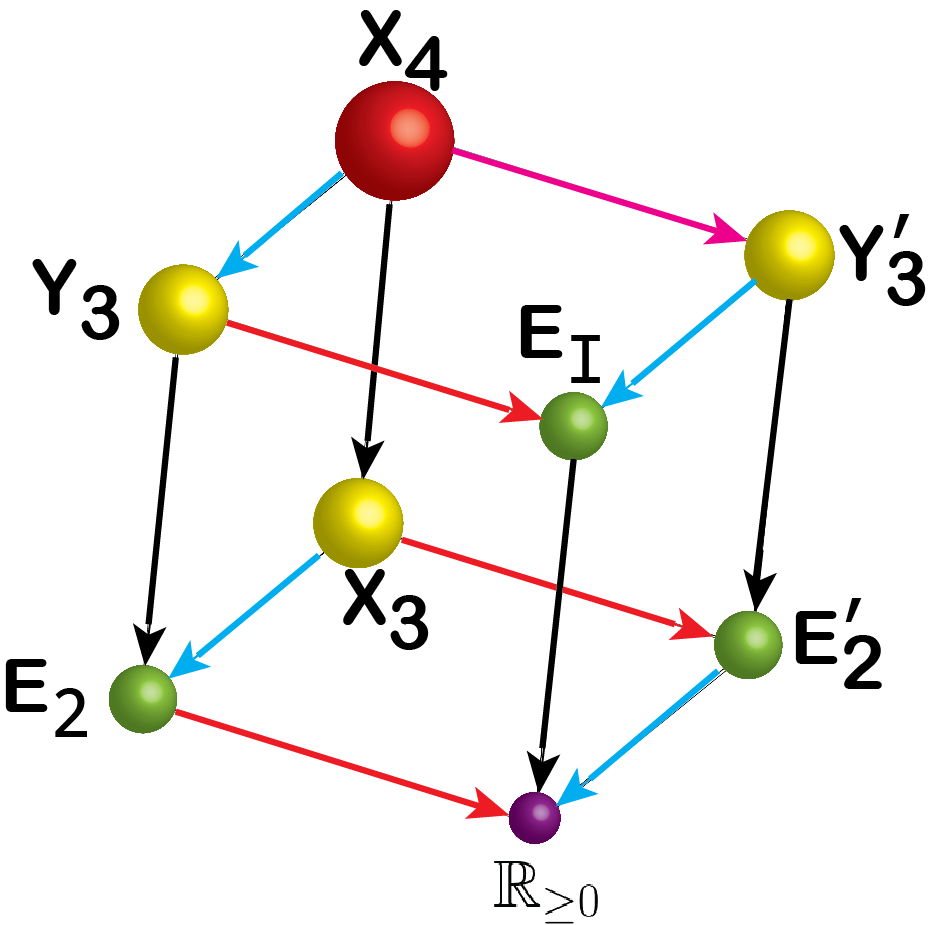}
		\caption{The figure shows a lattice whose vertices are various spaces that are underlying spaces
			$X$	for a possible phase space $(X,\A,B)$ for friction. The arrows are the forgetful projection maps (they omits certain information) between them.
			The coordinate description of these spaces are 
			$X_4 = \{(\norm{N}, \norm{F}/\norm{N}, \theta_{obj}, \phi)\}$, 
			$X_3= \{(\norm{F}/\norm{N}, \theta_{obj}, \phi)\}$,
			$Y_3= \{(\norm{N}, \norm{F}/\norm{N}, \theta_{obj})\}$,
			$Y_3'= \{(\norm{N}, \norm{F}/\norm{N}, \theta_{sub})\}$,
			$E_2= \{(\norm{F}/\norm{N}, \theta_{obj})\}$,
			$E_2'= \{(\norm{F}/\norm{N}, \theta_{sub})\}$,
			$E_I= \{(\norm{N}, \norm{F}/\norm{N})\}$,
			$\mathbb{R}_{\ge 0}= \{\norm{F}/\norm{N}\}$.	
			The four maps shown by black arrows leave out the magnitude  $\norm{N}$ of the normal force. The four blue maps forget $\theta_{sub}$ by leaving out the coordinate $\phi$ or the coordinate
			$\theta_{sub}$ as the case may be. The four parallel red maps 
			forget $\theta_{obj}$ by leaving out the coordinate $\theta_{obj}$
			or by replacing  the pair of numbers $(\theta_{obj},\,\phi)$ by the single number 
			$\theta_{sub} = \theta_{obj}+\phi$, 
			as the case may be.}
		\label{fig:latticespace}
	\end{figure}

A large class of many body systems ranging from  dense colloidal suspensions, granular matter to vortex matter in superconductors makes dynamical transitions between stuck and moving states
under the influence of an applied force  \cite{miguel2006jamming,blatter1994vortices,van2009jamming,reichhardt2016depinning}. 
The  transition that takes the system from a stuck to a moving state is usually referred to as 
{\it yielding} or {\it depinning} and this is associated with microscopic plastic processes
\cite{miguel2006jamming}. Similarly, the transition that takes a system from the moving (flowing) to a static state is referred to as 
{\it jamming}, and is marked by a drop in the single particle mobility \cite{kumar2013weak}. 
The problem of the motion of a single object has been studied earlier
from the point of view of the underlying microscopic mechanism of elastic instability \cite{tomlinson1929cvi,mcclelland1992friction,weiss1996dry,caroli1996dry} and plasticity
 \cite{bowden2001friction,greenwood1966contact}.  
In this paper we re-visit the phenomenology of onset and cessation of motion of a single object which is frictionally coupled to a substrate, and geometrically classify the various phases 
associated with it and the transitions between them, 
in terms of subregions of appropriately chosen phase spaces.

	Suppose that a stationary object is pressed against a homogeneous substrate by a normal force $N$, and a tangential force $F$ is then applied to the object (see Fig.\ref{fig:velo_gap}(a)).
	%, and we are interested in what happens next. 
	Depending on the force of static friction, 
	such an object will either remain stuck, or begin to move under the applied 
	force $F$. Correspondingly,
	we can say that the object is in a {\it stuck phase} or in a {\it slip phase}.
If we keep all other parameters constant, the maximum value of $|F|/|N|$ for which
the object remains fixed is called the coefficient of static friction, denoted by $\mu_{stat}$.
Suppose now a 
tangent force $F$ is applied to an object as above, which drives the object at a steady
velocity $v$. It is known that for moderate values $v$ and of $|N|$, the ratio $|F|/|N|$ 
is independent of $|N|$. 
The plot of $|F|/|N|$ against the velocity $v$ continues as the well known
Stribeck curve (the red curve in Fig. \ref{fig:velo_gap}  is a schematic representation of the 
$|F|/|N|$ vs. $v$ plot) \cite{woydt2010history}. The Stribeck curve
has portions of negative slope, and these are regions of instability. For if 
the slope is negative at a point $(v,|F|/|N|)$, and the value of $|F|/|N|$ is kept steady, 
then 
any slight increase in velocity will lead to an acceleration, and 
any slight decrease in the velocity will lead to a deceleration,
which will get further enhanced as long as we are in the part of the curve with negative slope. 

Note that by definition of $\mu_{stat}$, the vertical portion of the $|F|/|N|$ axis up to the 
point $(0,\mu_{stat})$ lies on the Stribeck curve. 
The Stribeck curve has a horizontal portion (where $|F|/|N|$ is independent of $v$) 
starting from a small non-zero velocity $v_0$. 
We call $v_0$ as the {\it velocity gap}. 
The corresponding value of $|F|/|N|$ is known as the coefficient of dynamic friction,
and it is denoted by $\mu_{dyn}$. 
It is known that $\mu_{dyn} < \mu_{stat}$, and 
for $0<v<v_0$, the Stribeck curve does not
have any noticable portions of continuity with positive or zero slopes.
The inequality  $\mu_{dyn} < \mu_{stat}$
means that if we apply a small mechanical disturbance to an
object in the stuck phase ($v=0$ and $|F|/|N| < \mu_{stat}$) 
to momentarily dislodge it, so as to allow the object 
to begin moving under the applied force $F$ with a speed $v_1 >v_0$, 
then there are two possibilities. For $|F|/|N|< \mu_{dyn}$, the object rapidly comes to
a halt again, while for $|F|/|N| > \mu_{dyn}$, the object continues to move. 
As $\mu_{dyn} < \mu_{stat}$, the portion of the Stribeck curve which lies over
$0< v< v_0$ must have portions of negative slope or discontinuities. The resulting unstable
behaviour of the object makes a precise plot of the Stribeck curve difficult in this region. 
The presence of discontinuities and regions of negative slope above $(0,v_0)$ 
mean that if in the above we have $\mu_{dyn}< |F|/|N|< \mu_{stat}$ and $v_1 < v_0$, then the object either rapidly comes to a halt 
or speeds up, or shows an intermittent behaviour mixing the above two possibilities.

%The velocity $v_0$, which we shall call as the {\it velocity gap}, is the value of $v$
%at which the Stribeck curve for the $1$-dimensional motion in the 
%direction of $F$ levels off within the region of boundary lubrication (see Fig.\ref{fig:velo_gap}). 

%
%If we apply a small mechanical disturbance to an
%object in the stuck phase (therefore $|F|/|N| < \mu_{stat}$) 
%to momentarily dislodge it, so as to allow the object 
%to begin moving under the applied force $F$ with a speed $v$ greater
%than a certain minimal speed $v_0$, 
%then there are two possibilities. For small values of $|F|/|N|$, the object rapidly comes to
%a halt again, while for higher values of $|F|/|N|$, the object continues to move. 
%The velocity $v_0$, which we shall call as the {\it velocity gap}, is the value of $v$
%at which the Stribeck curve for the $1$-dimensional motion in the 
%direction of $F$ levels off within the region of boundary lubrication (see Fig.\ref{fig:velo_gap}). 

%, transitioning from a region
%of dry friction to a region of mixed lubrication (see Fig.? in [blah]). 
%After the nearly vertical region, it crosses over
%to a region dominated by hydrodynamic lubrication (see blah). 
% In the Coulomb regime, with isotropic object and substrate,
% the critical value of $|F|/|N|$ where the  transition takes place is called 
% the coeficient of dynamic friction, and it is denoted by $\mu_{dyn}$

We will say that a stuck object is in a {\it strongly stuck phase} or a {\it weakly stuck phase}, depending respectively on whether the object comes back to a halt or continues moving after being 
subjected to a momentary disturbance with velocity $v_{1}> v_0$ as above.
As the Stribeck curves becomes approximately horizontal for a small interval after $v_0$ \cite{rabinowicz1958intrinsic}
there is a comfortable margin for $v_1$, which make the definitions of these phases robust. 
If the disturbance given to a weakly stuck object is so small as to give it
a velocity $v_1<v_0$, then the response will be intermittent 

 The above discussion was focussed on a $1$-dimension situation, where a given object is
	moving on a given homogeneous substrate by translational motion in a given direction. If the object or substrate
are not isotropic, the phases will depend on directional parameters, and the quantities
$\mu_{stat}$ and $\mu_{dyn}$ will not be constant but will depended on multiple parameters \cite{pabst2009anisotropic,yu2014understanding,yu2012friction}. 
	In this paper we study the geometric phenomenology of the above phases with multi-parametric 
	dependencies for a given pair of object and substrate with a given nature of contact.
	We set up an appropriate {\it phase space} $X$ together with nested subregions
	$X\supset \A\supset \B$, where points of $X$ parametrize the magnitudes of the forces on the stationary object and the angles between the force $F$ and fiducial directions on the object and the substrate, the subset $\A$ corresponds to those values for which the object is in a stuck phase, and the even smaller subset
	$\B$ corresponds to those values for which the object is in a strongly stuck phase. Note that 
	the set $X-\A$ corresponds to the slip phase, and $\A - \B$ corresponds to the weakly stuck phase.
	
	In the classical Coulomb regime of static friction, the phase regions $\A$ and $\B$ turn out to 
	determined by single numbers $\mu_{stat}$ (the coefficient of static friction) and $\mu_{dyn}$
	(the coefficient of dynamic friction). However, beyond the Coulomb regime, the regions $\A$ and $\B$ 
	become more complicated functions of the magnitude $|N|$ of the normal force and the angles between the force $F$ and fiducial directions on the object and the substrate. (We will leave out for simplicity the dependence of
	static friction on the contact time, as in \cite{sharma2008microrheology,ruina1983slip}.)
	We show that the shapes of these regions
	can depend on the macroscopic shape of the object, in particular, on its edges and corners. 
	
	The triples $(X,\A,\B)$, which are associated with a given object, substrate and a fixed nature of contact (this last notion is explained later), have a certain universal property. Namely, given 
	any experimental setting for friction, with a space $S$ of control parameters, there is a unique
	map $\psi$ from $S$ to $X$, such that the object with control parameters given by 
	a point $s\in S$ is in a certain phase if and only if its image $\psi(s)$ is in the corresponding subregion of $X$. The phase space $(X,\A,\B)$ for the given object, substrate and nature of contact
	is characterized by this universal property. 
	The map 
	$\psi: S\to X$ will be called as the {\it classifying map} for the experiment. We illustrate these concepts with some experiments.

	\subsection*{Phase spaces for friction\\ and their coordinate descriptions}

%		We recall that a modulus is a parameter which characterizes some quantity of interest. 
%	For example, the Young modulus characterizes the elasticity property of a material.
%	The word moduli is the plural form, used when more than one parameter is needed.
%	These parameters can be regarded as coordinates (sometimes only locally defined) on a 
%	space $X$, called the moduli space for the phenomenon in question.
%	If the phenomenon varies with control parameters which form a space $S$, then we get a 
%   classifying  
%	map $\psi: S\to X$ which captures this variation. In general, moduli spaces enjoy a certain 
%   `universal property', which we will discuss for static friction in the last section.

		\begin{figure}[b]
	\includegraphics[width=0.8\linewidth]{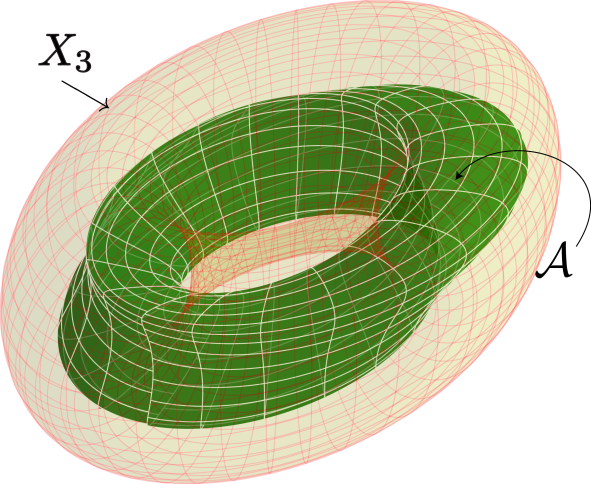}
	\caption{The yellowish solid torus depicts the space $X_3$, which is the phase space when both the object and the substrate are anisotropic, and when $\norm{N}$ is moderate. The darker greenish region is a schematic representation of the region $\A$ for a hypothetical object and substrate. }
	\label{Fig:torous}
\end{figure}

	For the basic experiment with static friction, let there be chosen 
	a pair of orthogonal directions on the object. We place the object on the substrate in any
	manner in which the first of these directions is normal to the substrate. The second chosen direction on the object, which is therefore tangent to the substrate, will be called as the fiducial direction on the object. Let there be also chosen 
	a fiducial direction on the substrate, which is tangent to its surface. Note that these two fiducial directions and the applied
	force $F$ are coplanar, as all three are tangent to the surface.
	The choice of the fiducial directions on the object and the substrate matter exactly when both the object and the substrate 
	are nonisotropic. In this case, let $\phi $ denote the angle from the fiducial direction on the object to the fiducial direction on the substrate. 
	Let $\theta_{obj}$ and $\theta_{sub}$ respectively denote the angle between the applied force $F$ and the 
	fiducial direction on the object or on the substrate. 
	Unlike $\phi$, the angles $\theta_{obj}$ and $\theta_{sub}$ are indeterminate when $F =0$.
	When $F\ne 0$, we have the equality
	$\phi  =  \theta_{sub}-\theta_{obj}$. We will always assume that $N$ is  
	non-zero, which is physically justified by the presence of gravity and 
	adhesion in our actual experiments, and so $\norm{N} \in \R^+ = \{ r\in \R\,|\, r>0\}$,
	the open positive half line. As $F$ is allowed to be zero, the ratio $\norm{F}/\norm{N}$ lies
	in the closed positive real half-line 
	$\R_{\ge 0} = \{ r\in \R\,|\, r \ge  0\}$. 
	As the coordinates $\theta_{obj}$ and $\phi $ are angular coordinates,
	their values correspond to points on the circle $S^1$.
	As the value of 
	$\theta_{obj}$ is indeterminate when $\norm{F}/\norm{N} =0$, that copy of $S^1$ gets squeezed to a point. 
	Hence, for a given value of $\norm{N} \in \R^+$, the pairs
	$(\norm{F}/\norm{N}, \theta_{obj})$ form a plane $E_2$ with polar coordinates $(\boldsymbol{r},\boldsymbol{\theta})$, where
	$\boldsymbol{r} = \norm{F}/\norm{N}$ and $\boldsymbol{\theta} = \theta_{obj}$. 
	Hence, the $4$-tuples 
	$(\norm{N}, \norm{F}/\norm{N}, \theta_{obj}, \phi )$ form the space $X_4 = \R^+ \times E_2 \times S^1$,
	which is a 
	$4$-manifold diffeomorphic to $ \R^3 \times S^1 $. 
	
	There are various simpler situations, where the object or the substrate or both are isotropic,
	or where we limit $\norm{N}$ to a range of moderate values, in which instead of the $4$-dimensional space 
	$X_4$, we can work with smaller $3$-dimensional spaces $Y_3$, $Y'_3$, $X_3$, or 
	$2$-dimensional spaces $E_I$, $E_2$, $E'_2$, or the half-line $\R_{\ge}$, as we now describe. All these spaces are sub-quotients of $X_4$. Their interrelationships are depicted as a cubical lattice of %quotient 
	spaces in Fig.~\ref{fig:latticespace}. The arrows stand for projection maps  which have a simple coordinate description in  coordinate terms.

	\noindent{\it The space $X_3$}: If we fix (or omit) the value of $\norm{N}$, then the triples $(\norm{F}/\norm{N}, \theta_{obj}, \phi)$ are points
	of $X_3 = E_2\times S^1$. Topologically, this space can be visualized as an open solid torus in $\R^3$.
	Alternatively, we can view it as $\R^3= E_2\times \R$ with the last coordinate being periodic with period $2\pi$, as the circle $S^1$ can be viewed as a line $\R$ with a periodic coordinate. 
	The Fig.~\ref{Fig:torous} schematically shows the way of visualizing $X_3$ as a solid torus, and depicts
	a subregion $\A$ within it. 
	
	\noindent{\it The spaces $Y_3$ and $Y'_3$}: If the object is anisotropic but the substrate
	is isotropic, then the angle $\theta_{sub}$ can be ignored, and so we get the space 
	$Y_3 = \R^+ \times E_2$ with coordinates $(\norm{N}, \norm{F}/\norm{N}, \theta_{obj})$,
	which is a $3$-manifold diffeomorphic to $ \R^3 $. Similarly, if the object is isotropic but the substrate
	is anisotropic, 
	%then the angle $\theta_{sub}$ can be ignored, and so 
	we get a space $Y'_3 = \R^+ \times E_2$, with coordinates $(\norm{N}, \norm{F}/\norm{N}, \theta_{sub})$ which is diffeomorphic to $Y_3$ with $\theta_{obj}=\theta_{sub}-\phi$.
	
	\noindent{\it The spaces $E_2$ and $E'_2$} : If we fix (or omit) $\norm{N}$, and if 
	the substrate is isotropic, so that we can ignore $\phi$ and $\theta{sub}$, 
	then the remaining data is in the form of pairs
	$(\norm{F}/\norm{N}, \theta_{obj})$, which gives a point (in polar coordinates) of the plane $E_2$.
	Similarly, if we fix (or omit) $\norm{N}$, and if 
	the object is isotropic, then we get pairs 
	$(\norm{F}/\norm{N}, \theta_{sub})$, which form a plane $E'_2$.  The spaces $E_2$ and $E'_2$ are both diffeomorphic to $\R^2$.
	
	\noindent{\it The space $E_I$}: If both the object and the substrate are isotropic, then the coordinates $\theta_{obj}$ and $\phi $ in $X_4$ can be ignored. In this case, we need to only consider the 
	pairs $(\norm{N}, \norm{F}/\norm{N})$, which form the $2$-dimensional space $E_I = \R^+ \times \R^{\ge 0}$, which is the first quadrant
	in $\R^2$ with the boundary $\norm{N} =0$ removed.

	\noindent{\it The half line $\R_{\ge 0}$} : If we fix (or omit) $\norm{N}$, and if both the object and the
	substrate are isotropic so that we can forget both $\theta_{sub}$ and $\phi$, then we are left with only a single non-negative real number $\norm{F}/\norm{N}$,
	which is a point of the closed half line $\R_{\ge 0}$. In geometric terms, this is the 
	conventional Coulomb scenario in which the frictional response is characterized by  a single number.

	In coordinate terms, the maps $X_4\to X_3$,  $X_4 \to Y_3$ and $X_4\to Y_3'$ in the commutative diagram in Fig.~\ref{fig:latticespace}  respectively send the $4$-tuple $(\norm{N}, \norm{F}/\norm{N}, \theta_{obj})$  to the $3$-tuples $(\norm{F}/\norm{N}, \theta_{obj}, \phi)$, $(\norm{N}, \norm{F}/\norm{N}, \theta_{obj}) $ and  $(\norm{N}, \norm{F}/\norm{N}, \theta_{obj}+\phi)$. The other arrows have similar obvious coordinate descriptions.
	
	The classic experiment shows that there exists a region $\A$ within $X_4$, such that if $(\norm{N}, \norm{F}/\norm{N}, \theta_{obj}, \phi) \in \A$ then the 
	object remains stationary. We will call $\A$ as the {\it sticky region}. This geometric representation raises  a  natural question: what is the shape the sticky region $\A$?
	In particular, one can ask:  how does the  shape of the  object  affect the shape of the region $\A$? 
	In this paper, we mainly explore the phenomenology of the relationship between these two shapes. 
	We will also consider experiments in which one or more of the object and substrate are isotropic,
	and in this cases we will replace $X_4$ by a smaller space $X_3$, $Y_3$, $Y'_3$, $E_2$, 
	$E'_2$, $E_I$ or $\R_{\ge 0}$ 
	as appropriate, and define a sticky region $\A$ in it similarly.

	\subsection*{The partition  $\A=\B\cup\C$ }
	
	It is known that the force of friction sharply decreases when a stationary object is set in motion
	(e.g., see p. 144 of \cite{persson2013sliding}).
	Consequently, a smaller applied force is sufficient to sustain motion, compared to what is needed to initiate motion. 
	This geometrically results in a partition of the sticky region $\A$ into two subregions $\B$ and $\C$,  as follows. Consider a point
	$P= (\norm{N}, \norm{F}/\norm{N}, \theta_{obj}, \phi)$ of $\A$. This physically
	corresponds to the situation where 
	steady forces $F$ and $N$ are applied 
	to the stationary object, which leave the object stationary. 
	To this stationary object, on which the steady forces $F$ and $N$ are
	acting as above, we give a burst of random mechanical vibrations which 
	momentarily dislodges the object so that it commences to move under the force $F$,
	acquiring a minimum velocity $v_0$ (the value of $ v_0 $ may depend on the directional parameters $\phi, \, \theta_{obj}$, but as the Stribeck curve flattens from $ v_0 $ for a small interval, the below definitions of $\B$ and $\C$ are robust). Alternatively, the burst of vibrations may be 
	administered to the substrate to set the object in motion.
	As explained in the beginning, now there are two possibilities. Either the object 
	continues to move for some time, in which case we say that $P\in \C$, or the object rapidly comes to a permanent halt
	in which case we say that $P\in \B$. 
	If the disturbance resulted in an initial velocity
	less than $v_0$, the object will rapidly come to a halt if in $\B$, while it
	will show intermittent behaviour if in $\C$. This fact allows a test for determining 
	whether a point is in $\B$ or $\C$ without knowing the value of $v_0$, however, as the 
	Stribeck curve is nearly flat for some velocity range from $v_0$. 
	% We do not need to know $v_0$ very precisely by flatness of Stribeck curve.
	This phenomenon, of the velocity rapidly going to zero once it
	drops below a critical value $v_0$, was described by the term {\it velocity gap} in \cite{ghosh2016curvilinear}.
	This partitions $\A$ into two subregions $\B$ and $\C$. In case we have a fixed or a moderate value of $\norm{N}$, or one or both of the object and the substrate are isotropic, 
	we replace $X_4$ by the appropriate lowest dimensional space among  
	$X_3$, $Y_3$, $E_2$ etc. shown in Fig.~\ref{fig:latticespace}, and define analogously the decomposition $\A = \B \cup \C$ within this smaller dimensional space.

	\begin{figure}[t]
		\centering
		\includegraphics[width=0.9\linewidth]{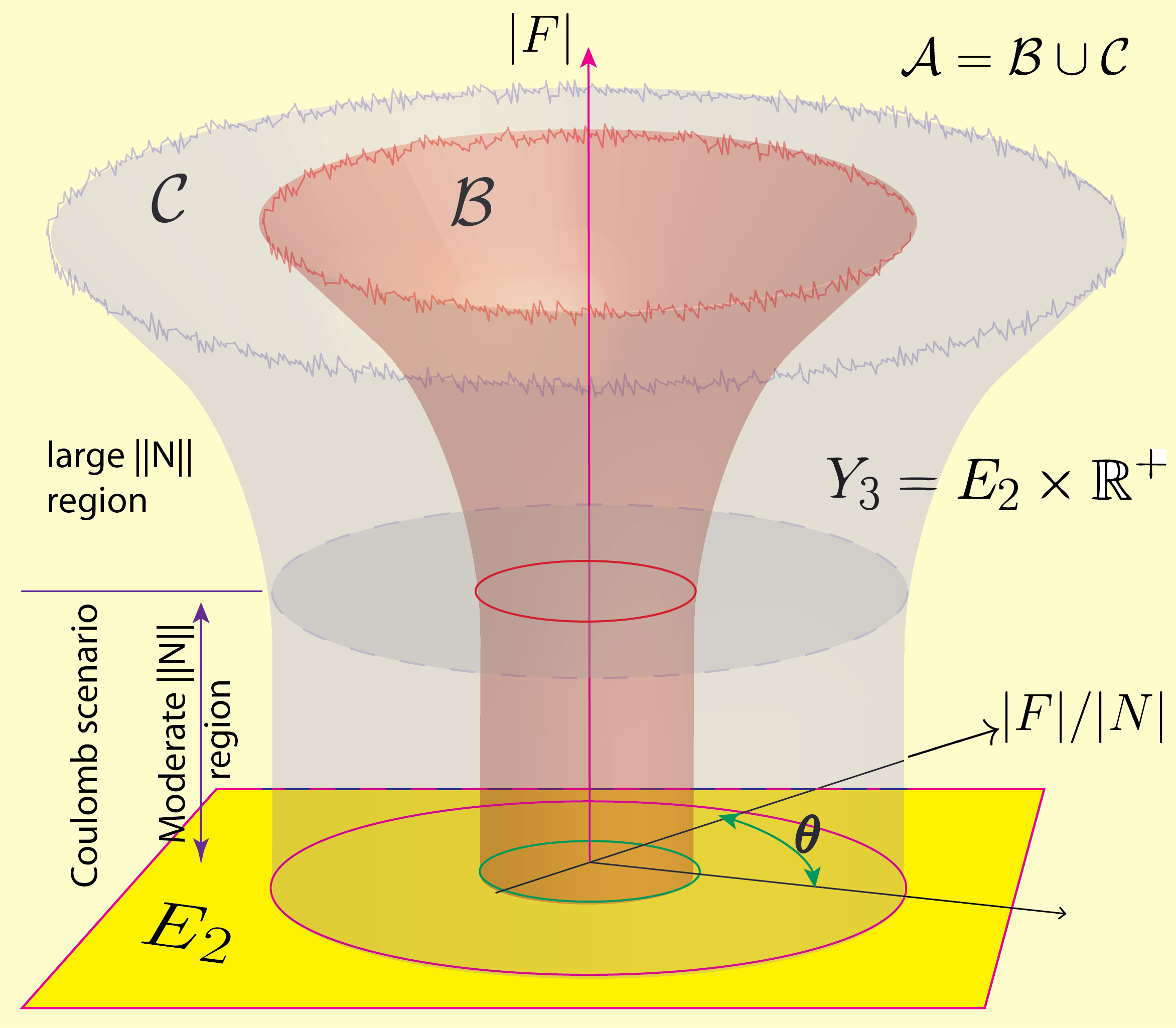}
		\caption{ The regions $\A$, $\B$  and $\C$ with variable $ \norm{N} $ in the space $Y_3=\R^+\times E_2$ with coordinates $ (\norm{N}, \norm{F}/\norm{N}, \theta_{sub}) $. As $ \norm{N} $ increases these regions  become broader.  For simplicity we have retained  rotational symmetry of these regions even for large  $ \norm{N} $.}
		\label{fig:schematic_onset}
	\end{figure}

The triple $(X_4, \A, \B)$ is the phase space for a frictional experiment, in a precise universal sense, as we explain later. In case the object or substrate are isotropic or $\norm{N}$ can be left out, 
the phase space will be 
the one of the  
triples $(X_3, \A, \B)$ etc., where the parameter space is the smallest, 
instead of $(X_4, \A, \B)$. When this happens, the regions $\A$ and $\B$ in $X_4$ etc. will be the inverse
image of the universal regions $\A$ and $\B$ in the phase space, under the projection arrows shown in Fig.~\ref{fig:latticespace}.

\subsection*{Diagrammatic representation of Coulomb friction}

	Phrased this way, the Coulomb laws for static friction 
	can be interpreted to say that  
	the region $\A$ is defined in $Y_3$ by a single inequality $\norm{F}/\norm{N} \le \mu_{stat}$
	in terms of a constant $\mu_{stat} \in \R^+$ which depends only on the  nature of the surfaces of 
	the object and the substrate.
	Similarly, the region $\B$ is defined in $Y_3$ by a single inequality $\norm{F}/\norm{N} \le \mu_{dyn}$
	where $\mu_{dyn} \in \R^+$ depends only on the nature of the surfaces of 
	the object and the substrate. The inequality $\mu_{dyn}< \mu_{stat}$ always holds.
	In particular, the shape of the object 
	does not affect the shapes of $\A$ and $\B$.
	This implies that the angle $\theta_{obj}$ is not relevant (for the original object placed with a 
	new angle $\theta_{obj}$ can be regarded as a new object with the same kind of surface). 
	These laws are known to be valid experimentally when $\norm{N}$ is not too large
	(what actually matters is that the resulting maximum pressure exerted at any point of the 
	interface is not too large, see pp 19-20 of  Ref. \cite{bowden2001friction}). 
	
	The Fig.~\ref{fig:schematic_onset} schematically depicts  the regions $\A$ and $\B$ in $Y_3$ in the Coulomb scenario. 
	These look like coaxial solid circular cylinders with radii
	$\mu_{dyn}< \mu_{stat}$ for small $\norm{N}$. Thus, for the Coulomb scenario, the phase space
	$(X,\A,\B)$ is the triple where $X = \R_{\ge 0}$, $\A =\{0\le r\le \mu_{stat}\}$ and $\B =\{
	0\le r \le \mu_{dyn}\}$.
		When  $\norm{N}$ becomes very large, we move out of the Coulomb scenario, i.e., the ratio $ \norm{F}/\norm{N} $ is no longer constant. Under the influence of a large normal force, the object begins to adhere to the substrate, which increases the radii of both $\A$ and $\B$ with $\norm{N}$.  The upper part of Fig.~\ref{fig:schematic_onset} schematically shows such an increase, where for simplicity, we have retained the rotational symmetry of $\A$ and $\B$. 
	Assuming such a rotational symmetry indeed holds, we must take $X = E_I$, while the regions $\A$ and $\B$ will have to be experimentally determined.

	\subsection*{A thought experiment  beyond the Coulomb scenario}
	
	In the laboratory experiments that we discuss later in the paper, we take $\norm{N}$ to be in the moderate range, so that 
	the object does not adhere to the surface and the frictional force is linear in $\norm{N}$. Hence in this range, we can ignore
	the coordinate $\norm{N}$, and just retain $\norm{F}/\norm{N}$. Assuming moreover that the substrate
	is isotropic, so that $\phi$ and $\theta_{sub}$ can be ignored,
	we can work in the $2$-dimensional  
	space $E_2$ with polar coordinates $(\norm{F}/\norm{N}, \theta_{obj})$, and take
	our regions $\A$ and $\B$ to be defined inside this space $E_2$, which was introduced earlier. 
		
\begin{figure}[t]
		\centering
		\includegraphics[width=1\linewidth]{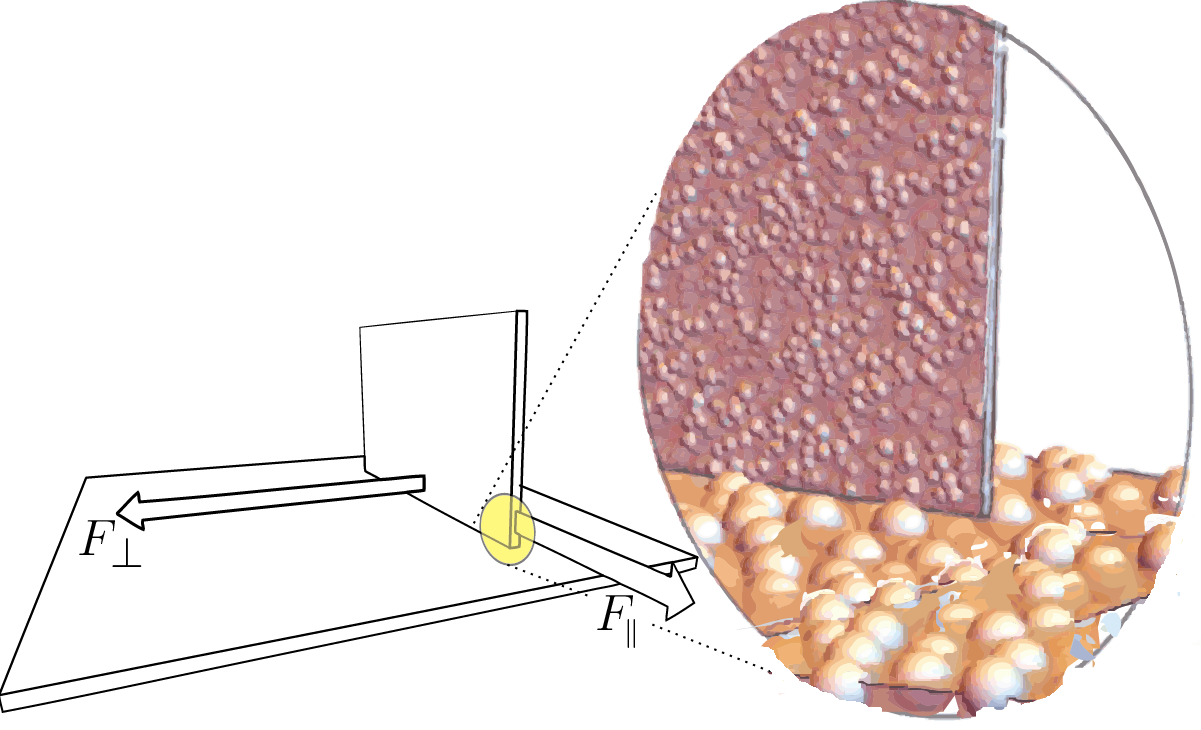}
		\caption{The figure shows a schematic representation of a thought experiment described in the text using cobblestone materials.}
		\label{fig:fig1cobblestone}
	\end{figure}

	In contrast to the Coulomb scenario, 
	we will show that when an object
	is placed on a substrate which is significantly rougher than the object, then
	the regions $\A$ and $\B$ may depend on the shape of the object. 
	Conversely, if the object is rougher and the 
	substrate is smoother, then this effect disappears.
	Before giving experimental data which shows this effect,  
	we present a simple idealization of the object and the substrate,
	for which one may see how the regions $\A$ and $\B$ are influenced by the macroscopic shape of the object.
	In this idealization, both the object and the substrate are fashioned from
	cobblestone material \cite{homola1990fundamental}. This kind of material is
	made by embedding rounded cobblestones of 
	nearly identical sizes in an elastic material. At the exposed surface of this material,
	we assume that at least half of each exposed cobblestone is embedded in the elastic
	material, which holds the cobblestones together. 
	We assume that the typical distance of closest approach between 
	two adjacent cobblestones is approximately the same as the diameter of a cobblestone. This distance $\ell$ defines
	a characteristic length scale associated with the object. 
	The magnified view in  Fig.~\ref{fig:fig1cobblestone} shows sheets made from two such materials which have different values of $\ell$.
	More complicated models of cobblestone materials will have a denser packing, or will have the interstices
	filled with smaller stones and so on, and there will be multiple length scales associated with the 
	material,
	but for simplicity, we will work with our model material which has a single length scale $\ell$.

	We assume that the material of the cobblestone
	as well as the elastic material have  
	a common non-zero coefficient of static friction against any of these two materials. Assuming different values for the
	possible coefficients, though more realistic, will not make a qualitative difference to the outcome
	of the following.
	
	We now describe a 
	thought-experiment in which an object in the form of a thin sheet fashioned from cobblestone material 
	with length scale $\ell_{obj}$ is placed on a substrate fashioned from cobblestone material 
	with length scale $\ell_{sub}$, such that $\ell_{obj}$ and 
	$\ell_{sub}$ are significantly different, with $\ell_{obj} \ll \ell_{sub}$. 
	The object is assumed to be thin in the sense that the thickness of the sheet from which it is fashioned is significantly smaller than the characteristic length scale $\ell_{sub}$ associated with the substrate. 
	However, the breadth and length of the object (unlike the thickness of the object) are assumed to be
	orders of magnitude larger than the sizes of the cobblestones. 
	This is depicted in the  magnified view  within  Fig.~\ref{fig:fig1cobblestone}, with the thin sheet perpendicular to the substrate. It can be seen that the vertical thin sheet
	gets partially embedded in the gaps of the cobblestones of the substrate. For this to happen, the elasticity of the sheet comes into play, allowing it to bend slightly to fit better 
	in the valleys between the protruding cobblestones of the substrate (in contrast, a completely  rigid  sheet  will rest itself on just  a few of the cobblestones and the edge will not get embedded into the gaps  between the cobblestones of the substrate).  In order to move laterally under the force $F_{\perp}$,
	the sheet has to climb over each of a larger number of such protrusions to begin its motion, compared
	with what it has to do in order to move longitudinally under the force $F_{\|}$. The thinness and the 
	elasticity of the sheet, which allows it to bend, comes into play when it moves 
	longitudinally, allowing it to navigate by threading through the gaps between the cobblestones, so that it needs to climb only over a smaller number of them. Our assumption, that the 
	thickness of the sheet from which the object is fashioned is significantly smaller than the characteristic length scale $\ell_{sub}$ associated with the substrate, is crucial here.
	If the rectangular sheet has its corner clipped, so as
	to present a rounded corner, with its radius of curvature greater than  
	the characteristic length $\ell_{sub}$ of the substrate, then the longitudinal
	motion in the direction of the rounded corner becomes even easier, compared with 
	longitudinal motion when the corner is sharp. We call the phenomenon of extra
	frictional response due to a leading sharp corner of the object as the {\it corner effect}.
	So long as there is a leading corner, such an effect will show itself even if the sheet is not perpendicular to the substrate.
		The difficulty of moving laterally increases with the 
	length of the sheet, as its edge has to climb over a greater number of cobblestones on the substrate. We call this phenomenon as the {\it edge effect}.

	For the edge effect to manifest itself, the edge should be sharp at a scale 
	of the order of the scale of granularity of the substrate -- a more rounded `edge' will not
	show this effect. As a result of the edge effect,
	the region $\A$ becomes broadened in the direction perpendicular to the edge, to an extent 
	which depends on the length of the edge.

\begin{figure*}[t]
		\begin{center}
	%		{\includegraphics[width=0.95\linewidth]{python/paper_expts1.pdf}}
     		{\includegraphics[width=0.85\linewidth]{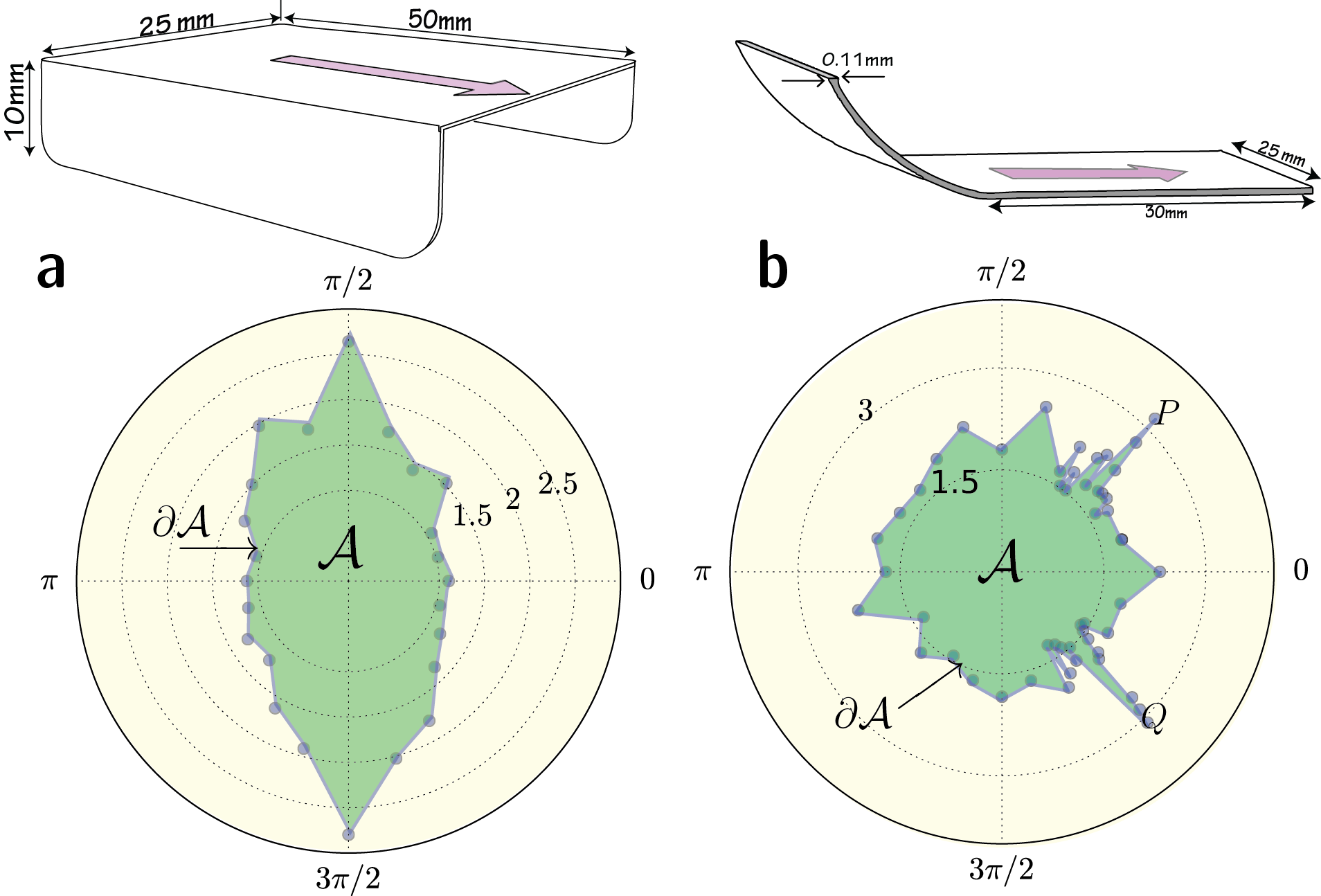}}
				\end{center} 
		\caption{The figure shows the measured effects of edges and corners on the region 
			$\A$ in two experiments, in which a paper sledge and a curled paper sheet
			were placed on a grit paper substrate.
			We used $ 100\,\mu\mbox{m}$ thick bond paper for fashioning both the sledge and the curled sheet, and used a P180 silicon carbide  grit paper with 
			average spacing between grit particles of the order of $ 100\,\mu \mathrm{m} $
			as the substrate. `The fiducial vectors on the objects are depicted as
			arrows drawn on them. The sledge in (a), made by bending a paper sheet,
			has rounded corers on its blades to lessen the corner effect.
			The blades give rise to a prominent edge effect, as the edges are vertical to the 
			substrate. The paper sheet in (b) is curled
			at one end, so that it has corners only on one side. The edge effect for a horizontally
			resting 
			sheet is not strong as its edge does not get much embedded into the substrate.  
			The points $P$ and $Q$ demonstrate existence of the corner effect,
	due the physical corners at either end of the leading edge of the curled paper. 
	In both these figures, because of the presence of
	noise, the data points are shown to be just {\it inside} the schematic plots of the corresponding regions.
	 }		
 \label{fig:sledge}
	\end{figure*}

	Instead of the sheet being perpendicular to the substrate, we can consider an 
	arrangement where it has an acute angle $\alpha$. 
	In this case, the motion of the sheet in the forward direction  
	will be more difficult than its motion in the opposite direction. 
	An extreme example of this is 
	when the angle $\alpha$ is $0$, that is, the object is a sheet is lying on the substrate. To isolate
	the effect of the leading edge, we will assume that the object is fashioned out of a rectangular sheet, and the opposite edge is raised by curling upwards
	(see Fig.~\ref{fig:sledge}(a), and also Fig.~\ref{fig:fig2}(b)). Such an object will have have a smaller friction moving in the direction of the 
	curled edge, and much higher friction moving in the direction of the straight edge.

The nature of the contact between the leading edge of the object and the substrate in this experiment will depend on the extent of the rigidity of the object. 
Given that the macroscopic dimension of the object is several orders of magnitude
greater than $\ell_{sub}$, the object will remain almost flat and its
leading
edge will remain almost straight, but it will get stuck against the most protruding of 
cobblestones from the substrate.  This will lead to greater resistance to move in the
direction of any leading straight edge.

	Though the actual materials involved are not cobblestone materials, the idea behind the above thought experiment gets realized in the experiment performed by placing a curled paper
	sheet on a rougher substrate, described later.
	
There is another scenario where a similar effect of the edge becomes manifest. This involves 
a block placed on substrate made of cobblestone material such that the substrate is elastic,
and the block is smoother than the substrate. The substrate develops a depression because
of the force $N$ on the object, and once again, any motion of the block 
transverse to an edge of it, or in the direction of a sharp corner of it, encounters the protruding cobblestones of the substrate as obstructions. This leads again to an edge effect or
a corner effect on the resulting regions $\A$ and $\B$, which get broadened in the direction 
perpendicular to an edge or in the direction of a corner. 
	Consequently, for an object with prominent corners and edges,
	the shape of the region $\A$, instead of being a circular disk in $E_2$ centred at the origin as in the Coulomb scenario, will
	come to depend on the shape of the object. For an object with no sharp corners, if the 
	ratio 
	$$ \frac{\mbox{edge length} \times \ell_{sub}}{\mbox{area of contact}}
	$$  goes to zero, then the Coulomb scenario will get restored. These edge effect scuffs  the  material  and produces  scratches on the  surface \cite{nilsson2009effects}.   The corner effect  is  accompanied by large stress concentrations, as a  result  the contacting objects  dig into one another  which produces tear \cite{klemenz2014atomic}.

\subsection*{The laboratory experiments}

		We can now say that the classic table top experiment will produce a constant $\mu_{stat}$,
		that is, a region $\A$ as in the lower (moderate $|N|$) part of Fig.~\ref{fig:schematic_onset}, 
under the following assumptions:
(a) the surface of the object is made of an isotropic material,
	(b) the surface of the substrate is made of an isotropic material, 
	(c) the edge effects are negligible, (d) the corner effects are negligible,  
	(e) the  toppling torque produced by the force $F$ is negligible (this is achieved if the point of application of $F$ is low enough
	-- a counterexample is given by rolling motion) and (f) the magnitude of $N$ is not too large.
	
	We are interested in finding the frictional response, more precisely, in finding the shapes of
	the regions $\A$ and $\B$, when some of the above assumptions (except (f),
	which we will not violate) are transcended. As we are leaving out the magnitude $\norm{N}$
	(and retaining just $\norm{F}/\norm{N}$), we can work in one of the spaces $X_3$, $E_2$, $E_2'$ or $R_{\ge 0}$ according to the below table,
	and define the regions $\A$ and $\B$ in this space.  
	
\begin{table}[h]
	\centering
	\label{table1}
	\begin{tabular}{c|c|c|}
		\cline{2-3}
		& \begin{tabular}[c]{@{}l@{}}Anisotropic\\   Substrate\end{tabular} & \begin{tabular}[c]{@{}l@{}}Isotropic \\ Substrate\end{tabular} \\ \hline
		\multicolumn{1}{|l|}{\begin{tabular}[c]{@{}l@{}}Anisotropic\\  Object\end{tabular}} & $X_3$                                                             & $E_2$                                                          \\ \hline
		\multicolumn{1}{|l|}{\begin{tabular}[c]{@{}l@{}}Isotropic\\  Object\end{tabular}}   & $E_2'$                                                            & $\R_{\ge0}$                                                    \\ \hline
	\end{tabular}
%\caption*{}
\end{table}

	For this, we use two different experimental arrangements, which respectively use as the substrate an inclined plane or the inner surface of a slowly rotating hollow horizontal cylinder.

	\begin{figure*}[t]
		\centering
		\includegraphics[width=0.9\linewidth]{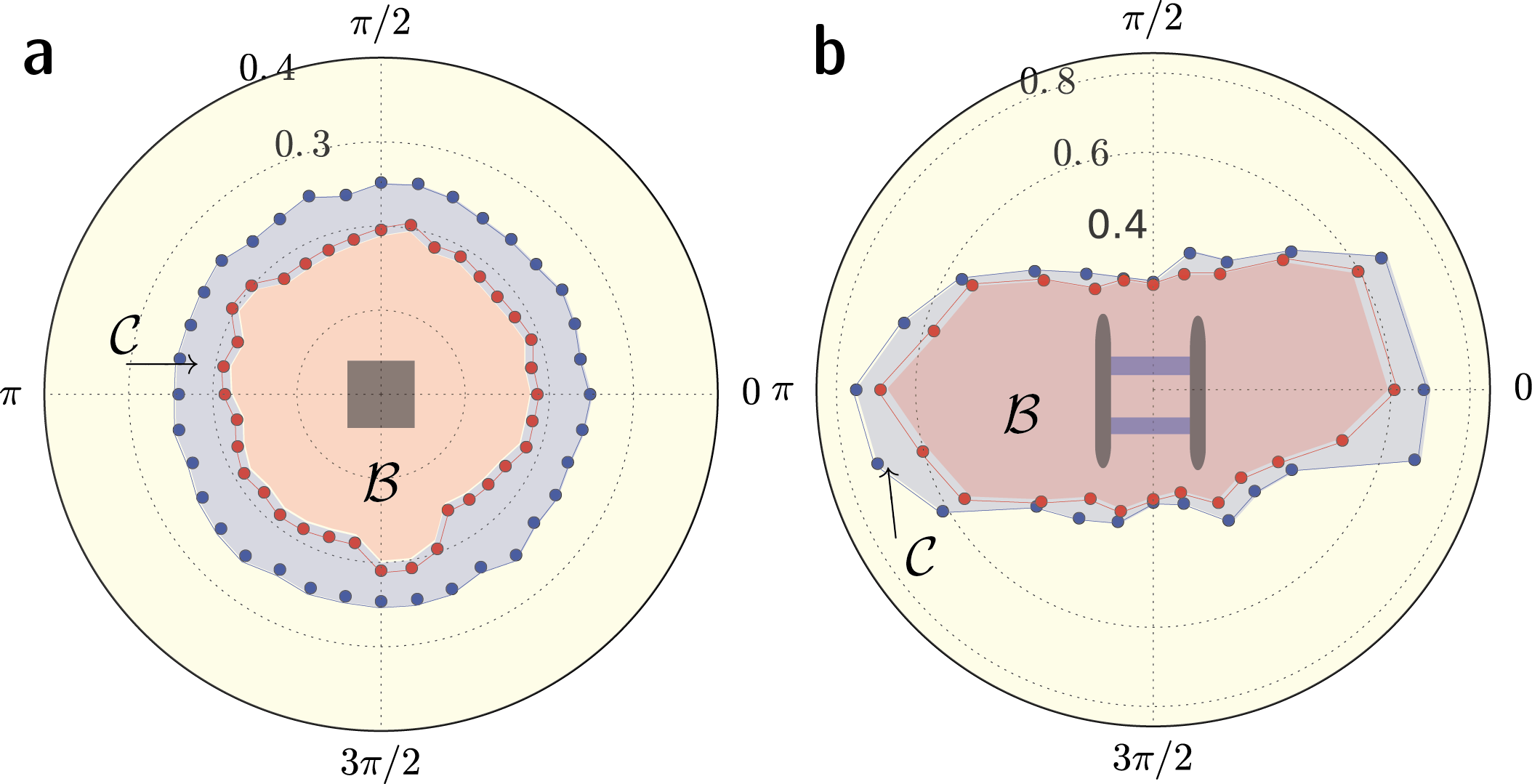}
		\caption{(a) The figure shows the empirically plotted regions $\C$ (blue) and $\B$ (red) of $\A(=\B\cup\C)$ for a square plastic  block 
			(sides $= 90 \, \mbox{mm}$, height $= 10 \, \mbox{mm} $). 
			In spite of the absence of circular symmetry in the block, 
			the regions are approximately circular,
			showing that corner and edge effects are negligible in this case. 
			(b) The figure shows the  motion onset curve $\partial \A$ (blue),the  subregion $\C$ (blue) and  $\B$  (red) of $\A(=\B\cup\C)$ for a sledge placed on a neoprene sheet. The sledge is made by gluing surgical blades to opposite sides of a square plastic block (length $= 90 \, \mbox{mm}$, height $= 10 \, \mbox{mm} $).
		In both these figures, 
		the data points in blue are shown to be just {\it inside} the schematic plots of the corresponding regions, while the data points in red are just {\it outside} the region $\B$. }
		\label{fig:fig2}
	\end{figure*}

		\begin{figure*}[t]
		\centering
			\includegraphics[width=0.99\linewidth]{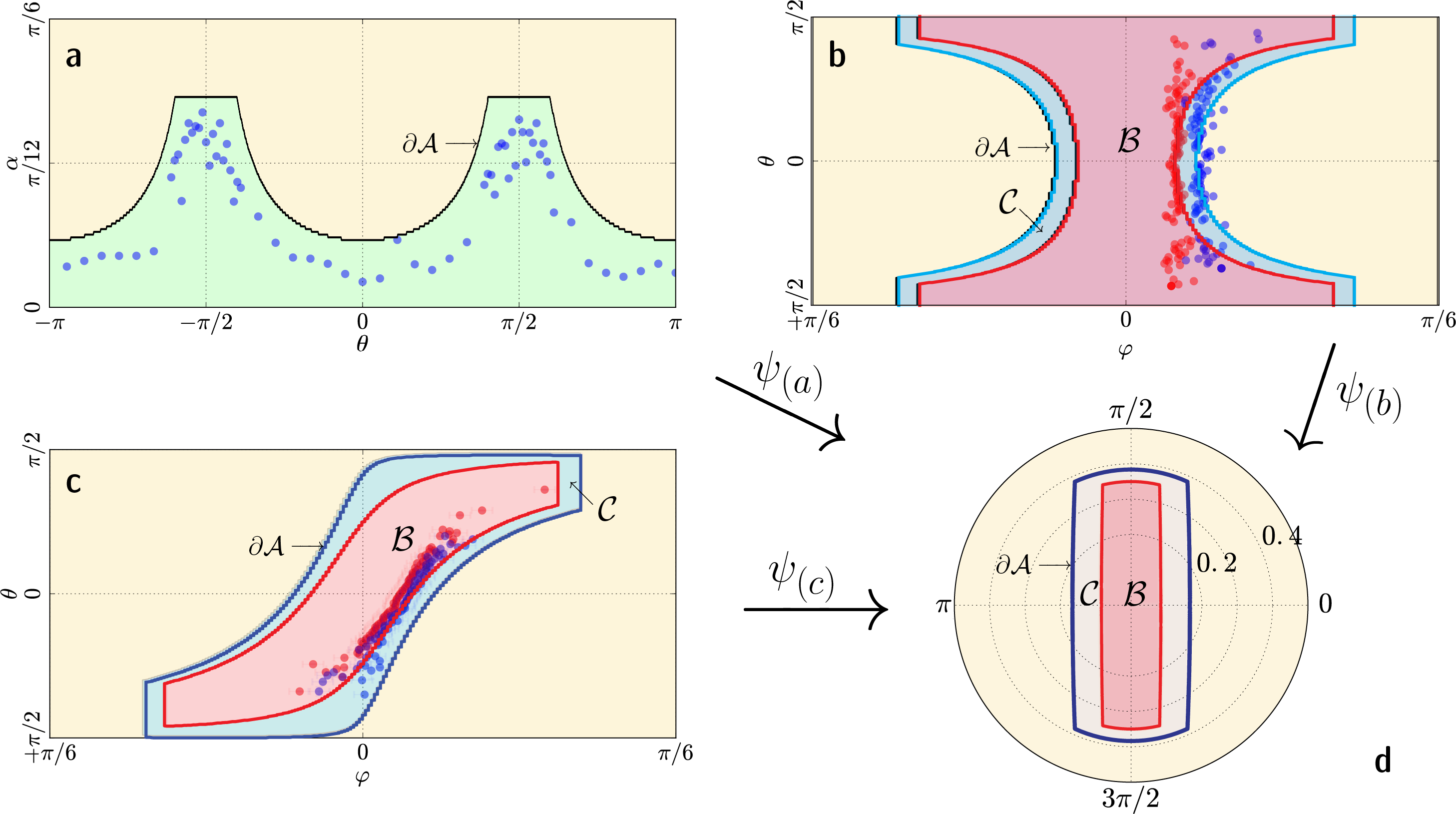}
		\caption{{\bf Classifying maps for frictional experiments with dumbbell}:  The panels (a), (b) and (c)  show the plots for the onset curved for a dumbbell in terms of the control parameters for three different experimental arrangements. These  onset curves are the pull backs  under the respective classifying maps 
	    $ \psi_{a} $, $ \psi_{b} $ and $ \psi_{c} $ of the universal onset curve in the phase space $E_2$ associated with a dumbbell on a glass substrate, shown in panel (d) with polar coordinates 
        $(\boldsymbol{r}, \boldsymbol{\theta})$.   
			\textit{Inclined plane apparatus}: The figure in panel (a) shows the region $\A$
			 plotted for a dumbbell placed on an inclined plane,  in terms of the control parameters $\alpha$  and $\theta$. The map $\psi_{(a)}$ is given by  $\boldsymbol{r}=\tan\alpha$ and $\boldsymbol{\theta}=\theta$. 
			 \textit{Horizontal cylinder apparatus}: The figure in panel (b) shows the  regions $\A$ and $\B$ plotted for a dumbbell placed on the inner surface of a horizontal cylinder apparatus,  in terms of the control parameters  $\varphi$ and $\theta$. The map $\psi_{(b)}$ is given by 
			 $\boldsymbol{r}=\tan\varphi$ and $\boldsymbol{\theta}=\theta$. 
			 \textit{Tilted cylinder apparatus}: The figure in panel (c) shows the regions $\A$ and $\B$ 
			 plotted for a dumbbell placed on the inner surface of a tilted cylinder apparatus
			 in terms of the control parameters $\varphi$ and $\theta$. The tilt of the cylinder
			 from the horizontal is fixed at $\alpha = 7^{\circ}$.
			 The map $\psi_{(c)}$ is given by $\boldsymbol{r}=\sqrt{(\cos\alpha\sin\varphi)^2+\sin^2\alpha}/(\cos\alpha\cos\varphi)$ and 		 
			 $\boldsymbol{\theta} = \arccos\left((\sin\varphi\cos\theta - \tan\alpha \sin\theta)/(|\sqrt{\sin^2 \varphi + \tan^2 \alpha}|)\right)$.
			 In (a), the plotted data points in blue are inside $\A$, 
			 while in (b) and (c), the blue points are inside $\A$ while the red points are inside $\B$.  To plot  $ \A $ we used $ \mu_r^{stat}=0.12 $ and $ \mu_s^{stat}=0.40 $. Similarly, to plot  $ \B $ we used $ \mu_r^{dyn}=0.08 $ and $ \mu_s^{dyn}=0.36 $.
			 }
		\label{fig:dumbbell_map}
	\end{figure*}

	When the substrate is an inclined plane, the forces $F$ and $N$ are
	supplied by gravity, with $\norm{F}/\norm{N} = \tan \alpha$ where $\alpha$ is the inclination of the plane. We assume that we are in the realm of moderate $|N|$ even when the plane is horizontal so that
	$|N|$ is at its maximum.
	We note that $\alpha = 0$ corresponds to a horizontal plane, and $\alpha = \pi/2$ corresponds to a vertical plane. 
	Therefore, $\theta_{obj}$ is the angle between the chosen fiducial vector on the object and the downward direction on the inclined plane. We vary $\theta_{obj}$ by rotating the object around an axis perpendicular to the substrate.
	The height of the object is kept small compared to its lateral dimensions whenever 
	we want the toppling torque produced by $F$ to be
	negligible. If one needs to test a non-isotropic substrate using this apparatus, this is possible
	by holding the angle $\theta_{sub}$ to be constant as $\alpha$ is slowly varied, and then repeating
	the experiment for a new value of $\theta_{sub}$.
		The angle of inclination  $\alpha$ of the plane is mechanically increased slowly,
	at the  rate of approximately $50$ degree per hour, starting with $\alpha =0$. 
	For different chosen values of $\theta_{obj}$, 
	note was made of the value of $\alpha$ at which the object began to move. 
	This gave various data points with polar coordinates $(r, \theta) = (\tan \alpha, \, \theta_{obj})$
	in $\A$, close to its boundary $\partial \A$ (the 
	onset curve), which enabled a schematic plot of the region $\A$. 
	
	This arrangement is used to plot the region $\A$ in $E_2$ (with coordinates
	$(\norm{F}/\norm{N}, \theta_{obj})$ as described above), for different objects. If the object placed on the inclined plane is a circular disk and moreover the substrate is isotropic, it is clear by symmetry that the region 
	$\A$ is the circular disk centred at the origin $r=0$ of $E_2$ with radius $\mu_{stat}$, 
	and the onset curve $\partial \A$ is its boundary circle $r = \mu_{stat}$. If the object is not circular, there is no a priori reason for
	the onset curve to be a circle as above, and the actual shape of it must be measured experimentally.

	The Figures \ref{fig:sledge} and \ref{fig:fig2}   show the observations and the plotted onset curves for 
	(1) a paper sledge on a grit paper,
	(2) a paper curled on one side  on  grit paper,
	(3) a plastic square block placed on a glass plate, 
	(4) a sledge with metallic blades placed on a neoprene sheet and
	(5) a dumbbell placed on an glass plate.
	To minimize the toppling torque on the non-rolling objects (cases (1) to (4) above), 
	their height was kept small compared to their lateral dimensions.
	
	To plot the region $\B$, the plane was first kept inclined at various fixed values of the 
	angle $\alpha$ at which the object remained stationary, and then a small burst of mechanical noise 
	was imparted to the assembly, and its effect was observed. 
	Note was made of whether the resulting motion of the object was transient or was a 
	sustained movement. These observations approximately gave the boundary 
	$\partial \B$ and so led to a schematic plot of region 
	$\B$  (see Fig.~\ref{fig:fig2}). This experiment was done for a square block and a metal sledge. 
	(The paper sledge and the plastic dumbbell, being too light, were not convenient for the administration of a mechanical burst, as they often got thrown off by the noise.)
	
	The second experimental arrangement, namely, a rotating hollow horizontal cylinder, was 
	used to plot the regions $\A$ and $\B$ and the onset curve $\partial \A$ for a dumbbell.  In this experiment  a hollow glass cylinder (radius$\,=125 \,\mbox{mm}$) whose axis is horizontal  is rotated very slowly around its axis (at an angular velocity $\approx 0.01 \, \mbox{radians/sec}$).
	A dumbbell made of plastic (radius of the balls$\,= 3 \,\mbox{mm}$, length of the  dumbbell$\,= 23 \, \mbox{mm} $) is placed at different angles 
	(values of $\theta_{obj}$) at the lowermost point on the interior surface of the cylinder, and its subsequent motion is observed. 
	As the radius of the cylinder is much larger than the size of the object placed on its surface,
	the surface can be regarded as approximately flat at the scale of the object. In any experiment 
	with the cylinder apparatus, care must be
	taken that the nature of contact between the object and the cylindrical surface is similar to that between the object and a flat surface, which is possible for rolling objects such as balls or
	dumbbells.
	
	Some of the advantages of using the cylinder apparatus  over using an inclined plane are the following.
(i) The slope of the cylinder continuously changes from $0$ to $\pm \infty$. This makes it possible to
	test the frictional behaviour of an object placed on the cylinder for all values of the slope of the substrate. 
(ii) At the lowest part of the horizontal cylinder (along the line $\varphi = 0$) any stationary
object defines a point in the region $\B$, as it remains stationary even after a small disturbance.
The slow rotation of the cylinder, which has a very low noise level, enables us to slowly transport (without imparting 
	a significant amount of linear momentum) 
	a small object placed at $\varphi = 0$ to a region with a higher slope while being stationary w.r.t. the substrate, from where it begins to move down.  This allows the determination of the  boundary $\partial\A$. 
(iii) As the object moves down, the slope of the substrate goes to zero, and the object comes to a stop.
    The point where it comes to rest is in the region $\B$, but not precisely on the 
	boundary $\partial\B$, because of the downward momentum of the moving object. 
	Similarly, the presence of the mechanical noise of rotation allows the object to penetrate a bit
	into $\B$ instead of stopping just at its boundary. These effects being stochastic,
	observing the points where it stops thus enables an approximate conservative
	determination of the boundary $\partial\B$ of region $\B$. So instead of a sharp boundary, as 
	expected from the existence of the velocity gap mentioned earlier, we get a fuzzy boundary
	in the experiment.
(iv) The rotation of the cylinder does not change the geometry of the experimental setup. 
	The gentle rotation allows us to
	put bounds on the regions $\A$ and $\B$ without resorting to a sudden
	burst of mechanical noise (as in the inclined plane experiment where it is more difficult to control
	and has a pronounced destabilizing effect on an object which can roll, such as a dumbbell).
	
	The  limitation of a cylindrical substrate is that while we can put balls or 
	dumbbells on it, where the nature of the contact is very similar to that for a flat substrate,
	we cannot put blocks with flat faces on a cylindrical substrate without drastically altering the 
	nature of the contact.  Given the advantages and disadvantages of each,  
	one or both the experimental set ups were deployed as were appropriate.

\subsection*{Results and discussion} 
	
\begin{figure}[t]
		\centering
				\includegraphics[width=1\linewidth]{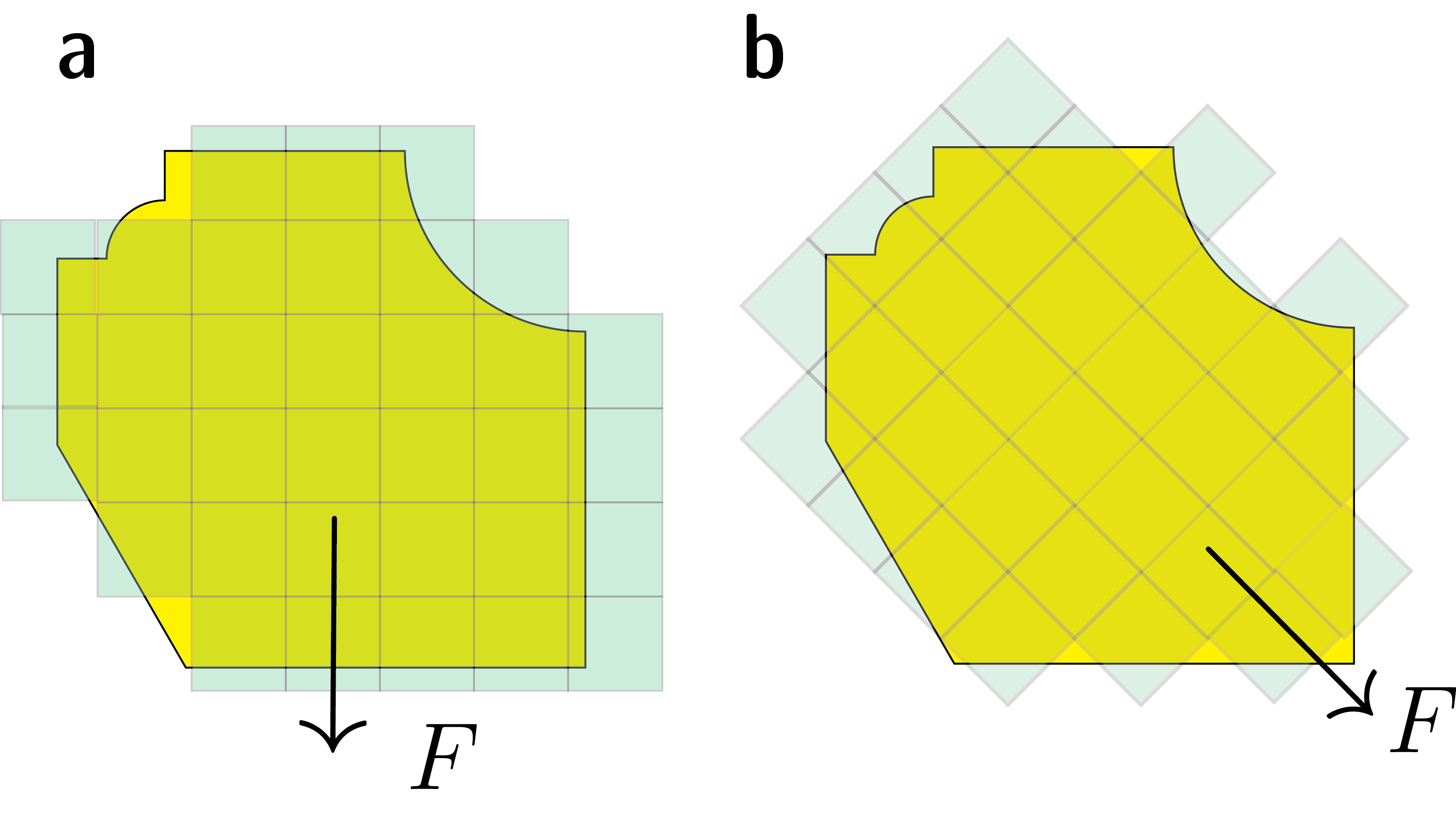}
		\caption{{\bf Isotropy of frictional response.}
			The figure show a flat, irregularly shaped, yellow 
			object, with piecewise smooth outer boundary, for which we assume that the  
			ratio of the area of the object to the perimeter of the object is small, and 
			whose boundary is not `too jagged'.
			A tangent force $F$ is applied in two different directions	
			in parts (a) and (b) of the figure. Let the object be approximated by a union 
			of squares joined together, oriented according to the direction of the applied force. 
			As the size of the squares is made arbitrarily small, the area of object gets approximated by the 
			total area of the squares to any desired precision. At the same time, the outer perimeter of the 
			assembly of squares approximates the perimeter of the block within a factor of $\sqrt{2}$ (a factor which comes from the worst case scenario of approximating an edge at an angle of $45^{\circ}$ to the direction of 
			$F$).
			Hence if the ratio of the area of the object to the perimeter of the object is small, then the 
			ratio of the total area of the assembly of squares to the  outer perimeter of the 
			assembly of squares	is again small. The imaginary edges of the squares which are glued to each other have
			no physical existence, and hence they have no effect on the friction. 
			This implies that the two assemblies of squares, and hence the object, 
			have the same frictional response in (a) and (b). 
		}
		\label{fig:figure3}
	\end{figure}

	\noindent
	1. {\it Square block in the Coulomb scenario.} When a square block 	is placed on an inclined plane, the experiment gives the expected results in the Coulomb scenario 
	%(namely, that $\A$ and $\B$ are concentric circular disks centred at the origin) 
	when the block has a low height (to minimize the toppling torque, which would have accentuated 
	the edge effect by putting extra pressure along the leading edge) 
	and has rounded edges or has a sufficiently large size (so that the ratio of the perimeter to area is small, making edge	effects negligible). 
	The regions $\A$ and $\B$ are measured to be concentric circular disks centred at the origin
	and the onset curve $\partial \A$ is a circle, in spite of the lack of 
	rotational symmetry in the shape of the block. In fact, this property of anisotropy of the 
	frictional response will hold for any shaped block 
	of large enough size whose perimeter is not too jagged, so that the ratio of edge length to
	area is small. This can be seen from Fig.~\ref{fig:figure3}, which explains why the frictional response remains 
	isotropic in terms of a reconstruction of the shape via conjoined square blocks which can be 
	chosen to have any given common orientation.

	\noindent
	2. {\it Curled paper sheet, and paper sledge, on a grit surface.}  The experiments described in Fig.~\ref{fig:sledge}
	are realizations of the thought experiments. 
	The  flat end of the paper shows greater resistance to onset of motion as compared to the curled edge, which is a manifestation of the edge effect, as expected. 
	As expected from the thought experiment, the paper sledge on the grit paper 
	moves most easily in the direction of its long axis, 
    and with greatest difficulty in the sideways direction.

\noindent
3. {\it Viscoelastic version of edge effect.}	
A variation on the above experiment is where the sledge is
	made by gluing surgical blades to the opposite sides of a plastic block is 
	placed on a deformable substrate made from neoprene (see Fig.~\ref{fig:fig2}(b)). These materials 
	are not in the domain of the thought experiment, which was with cobblestone materials.
However, deformations of a 
viscoelastic substrate and their propagation lead to corner and edge effects much like those discussed in the thought experiment. This follows from the dependence of the reaction force of a 
viscoelastic material on the deformation rate \cite{persson1998theory}, and the fact that the deformation is confined to a band around the edge with width equal to the characteristic length scale $s$ associated with the mechanical deformation of the body or the substrate. 
In this set up, the edge effect becomes prominent when the 
		ratio of the area
\begin{flushright}
	
\end{flushright}		of the contact region between the body and the substrate to the perimeter of the region of contact
		(this ratio has the dimension of length)	
		is of the smaller than the length scale $s$. Conversely, if the ratio 		
	$$ \frac{\mbox{edge length} \times s}{\mbox{area of contact}}$$  
	goes to zero, then the Coulomb scenario will get restored provided that the time rate of change of force is kept small (no sudden jerks are applied) to keep the viscous reaction negligible.
An edge effect of the above kind was indeed shown by the metallic sledge on the neoprene substrate
(see Fig.~\ref{fig:fig2}(b)).

	\noindent
	5. {\it Dumbbell.} 
	A dumbbell, made by joining two balls, can roll under a torque. 
	In this case, the regions $\A$ and $\B$ in $E_2$ 
%	which we plot using the cylinder apparatus, 
	are determined by four constants 
	$\mu^{stat}_s, \,\mu^{stat}_r,\,\mu^{dyn}_s, \,\mu^{dyn}_r$, 
	with $0 < \mu^{dyn}_r < \mu^{stat}_r \ll \mu^{dyn}_s < \mu^{stat}_s$. The definition and significance
	of these constants is explained in the Appendix A of \cite{ghosh2016curvilinear}.
The region $\A$ can be described as follows. Let $\epsilon^{stat} 
= \arcsin(\mu^{stat}_r/\mu^{stat}_s)$.
A point $(r,\theta)\in E_2$ with $\pi/2 - \epsilon \le  \theta_{obj}  \le \pi/2 + \epsilon$
or $\pi/2 + \epsilon \le  \theta_{obj}  \le  3\pi/2 -\epsilon$
lies in $\A$ if $r|\cos \theta| \le \mu^{stat}_r$. A  point $(r,\theta)\in E_2$ with 
$\pi/2 -\epsilon \le \theta_{obj} \le \pi/2 + \epsilon$ or 
$3\pi/2 -\epsilon \le \theta_{obj} \le 3\pi/2 + \epsilon$ 
lies in $\A$ if $r \le \mu^{stat}_s$. 
The theoretical expectation for the shape of the region $\B$ is given by replacing 
the static coefficients by their dynamic counterparts, and correspondingly replacing the 
constant $\epsilon^{stat}$. 
The experimental plots of $\A$ (made with an inclined plane as well as a rotating cylinder)
and the plot of $\B$ (made with a rotating cylinder), are consistent with the above
for moderate values of $\theta$.
On the other hand, a moving dumbbell for which $\theta$ is close to $\pm \pi/2$ (which therefore
slides more than it rolls) tends to change its value of $\theta_{obj}$ towards $0$ or $\pi$ as it slows down because of the torque produced by differential normal and tangent
forces on the two balls as well as due to mechanical irregularities. 
%This changes the value of $\theta_{obj}$ towards $0$ or $\pi$. 
	Consequently, our experimental plots of the regions
	$\A$ and $\B$ in Fig.~\ref{fig:dumbbell_map}
	are not closed when  $\theta$ is close to $\pm \pi/2$). 
	Noise (produced by the apparatus which raises the plane or rotates the cylinder) is constantly present in the experiments, so a dumbbell begins to move when it is somewhere within  
	the region $\C$, instead of starting from a point of $\partial \A$. The momentum of the moving dumbbell leads to its stopping somewhere inside the region $\B$, instead of stopping exactly on crossing $\partial \B$.

	\subsection*{Universal property of the phase spaces of friction}

	\begin{figure}[b]
		\centering
				\includegraphics[width=.75\linewidth]{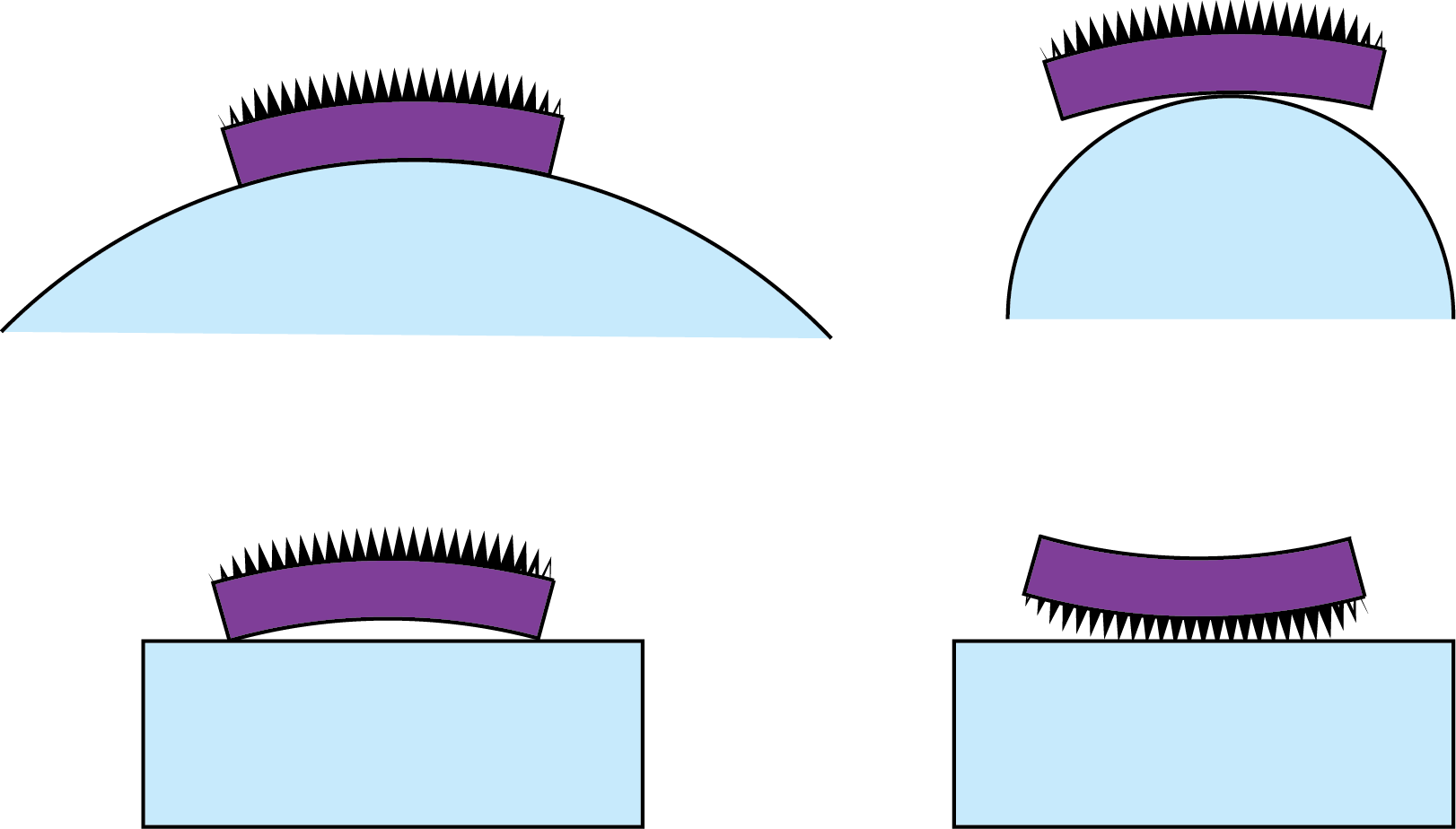}
		\caption{Examples of different natures of contact.
		The four panels of this figure schematically show an object $O$ coloured
		purple resting on different substrates made from  the same material $M$ coloured blue, such that 
	the nature of contact is different in each panel. Therefore, there will be four different
phase spaces $(X,\A,\B)$ for these arrangements, in which the underlying space $X$ will be the same
but the $\A$'s and $\B$'s will change. }
		\label{fig:natureofcontact}
	\end{figure}

	Consider any experimental arrangement, in which a chosen object $O$ is pressed against a surface 
	made with material $M$, with a normal force $N$ and subjected to a
	tangent force $F$, which may depend on the configuration of the
	object (its position and placement on the surface). 
	We do not need to assume that the surface is flat, but we must require that the 
	{\it nature of the contact} between the object and the surface
	is of a chosen sort
	(the Fig.~\ref{fig:natureofcontact} explicates with an example the concept of the `nature of the contact').
		Suppose that the surface is homogeneous, but not necessarily isotropic. If it is not isotropic,
	let a unit tangent vector field $V$ on the surface encode the possible directionality of the substrate (no such $V$ is to be given if the
	surface is isotropic). Let there be fixed a fiducial vector on the object. Then from the control parameter space $S$ of the object, we get a {\it classifying map} $\psi: S \to X_4$ to the phase space
	$X_4$ which sends a point of $S$ to the data $(\norm{N},\norm{F}/\norm{N}, \theta_{obj},\phi)$ for the configuration given by that point.
	If the surface is isotropic, we instead consider a map to $Y_3$.
	If we leave out $\norm{N}$ we get a map to $X_3$, etc. The phase spaces, together with their regions 
	$\A$ and $\B$, have the universal property that under the
	classifying map $\psi$, the inverse images 
	%$\psi^{-1}(\A)$ and $\psi^{-1}(\B)$ 
	of $\A$ and $\B$ 
	are the corresponding empirically determinable regions in the control parameter space $S$,
	where the object is in the stuck phase, and in the strongly stuck phase, respectively.

	Examples of the above for three different experiments with a dumbbell are given in 
	Fig.~\ref{fig:dumbbell_map}.	In other
	words, once we know by experimenting (say, by using an inclined plane) the 
	regions $\A$ and $\B$ in the phase space, the corresponding regions $\A$ and $\B$ in any other experimental set-up (whose control parameter space is $S$) can just be read off from the regions in the phase space by using the classifying map $\psi: S \to X$
		{\it without} once again performing the experiments in the new arrangement. Thus, we can get the regions $\A$ and $\B$ for any experimental arrangement from the {\it universal} regions $\A$ and $\B$ in phase space, by simply determining the classifying map to the appropriate
	 space. This is what is meant by the universal property of these spaces $X_4$ etc. as phase spaces of friction.
	
	When the object $O$ and the material $M$ (and the nature of contact) are altered, we 
	will get new regions $\A'$ and $\B'$ even if $X$ remains the same, giving a new phase space
	$(X,\A',\B')$. 
	
	It is possible to extend the construction of phase spaces of friction together with their
	universal subregions, which will have a universal property, 
	to experiments with composite objects, where the object has multiple parts which are joined by hinges, springs, etc. We can also increase the dimension of $X$ to incorporate the 
	effect of ageing of contact, as well as to incorporate a continuously variable nature 
	contact. 
	Also note that the classifying map $\psi: S\to X$ to the phase space $X$ preserves the symmetries of the experimental set-up, that is, symmetries of $S$, and so
	it descends to a map $\overline{\psi} : S/G \to X$ on the quotient space $S/G$ of $S$ by the group $G$ of all symmetries of the experimental set-up. This was used above for the 
	experiment with an inclined plane (where there is an action of $G =\R^2$ by translation) and
	for a cylinder (where there is an action of $G = \R$ by translation along the axis), and 
	what we have actually depicted in Fig.~\ref{fig:dumbbell_map} are the respective maps $\overline{\psi} : S/G \to X$.
	
%	\bibliography{onset} 
%merlin.mbs apsrev4-1.bst 2010-07-25 4.21a (PWD, AO, DPC) hacked
%Control: key (0)
%Control: author (8) initials jnrlst
%Control: editor formatted (1) identically to author
%Control: production of article title (-1) disabled
%Control: page (0) single
%Control: year (1) truncated
%Control: production of eprint (0) enabled
%

\end{document}